\def\be{\begin{equation}}
\def\ee{\end{equation}}
\def\ba{\begin{array}}
\def\ea{\end{array}}

\documentclass[prl,showpacs,twocolumn,amsmath]{revtex4}
\usepackage{graphicx}
\def\qed{\leavevmode\unskip\penalty9999 \hbox{}\nobreak\hfill
     \quad\hbox{\leavevmode  \hbox to.77778em{%
               \hfil\vrule   \vbox to.675em%
               {\hrule width.6em\vfil\hrule}\vrule\hfil}}
     \par\vskip3pt}

\newtheorem{theorem}{Theorem}
\newtheorem{corollary}{Corollary}

\begin{document}
\title{\large\bf Quantum separability criteria based on realignment moments}
\author{ Tinggui Zhang$^{1,2 \dag}$ Naihuan Jing$^{2,3}$ and Shao-Ming Fei$^{1,4,5}$}
\affiliation{ ${1}$ School of Mathematics and Statistics, Hainan Normal University, Haikou, 571158, China \\
$2$ College of Sciences, Shanghai University, Shanghai, 200444, China \\
$3$ Department of Mathematics, North Carolina State University\\
$4$ School of Mathematical Sciences, Capital Normal University, Beijing 100048, China \\
$5$ Max-Planck-Institute for Mathematics in the Sciences, Leipzig 04103, Germany\\
$^{\dag}$ Correspondence to tinggui333@163.com}
\bigskip

\begin{abstract}
{\bf Abstract}: Quantum entanglement is the core resource in quantum
information processing and quantum computing. It is an significant
challenge to effectively characterize the entanglement of quantum
states. Recently, elegant separability criterion is presented
in [Phys. Rev. Lett. 125, 200501 (2020)] by Elben \emph{et al.}
based on the first three partially transposed (PT) moments of
density matrices. Then in [Phys. Rev. Lett. 127, 060504 (2021)] Yu
\emph{et al.} proposed two general powerful criteria based on the PT
moments. In this paper, based on the realignment operations of matrices
we propose entanglement detection criteria in terms of such realignment moments.
We show by detailed example that the realignment moments can also be used
to identify quantum entanglement.


\end{abstract}

 \maketitle

\section{Introduction}
Quantum entanglement is one of the most profound features of quantum
mechanics \cite{rpkh}. It can be used to realize quantum tasks such
as quantum communication \cite{chgb,akek,cgcr}, quantum simulation
\cite{slio}, quantum computation and so on \cite{mail}. How to
characterize and quantify entanglement is one of the fundamental
problems in quantum information \cite{rpkh} and many-body quantum
physics \cite{jeis}. There has been a lot of efforts to explore the
quantum entanglement both theoretically and experimentally \cite{rpkh,oggt,nfgv}.

With the development of quantum information and quantum computing,
the research and development of medium-scale quantum devices
containing dozens or even more qubits is a hot topic
\cite{jpre,hszh,mgon}. For such devices, the standard methods of
tomography \cite{kvhr,mpjr} and witnesses \cite{oggt,nfgv} are no
longer feasible for gauging the performance in actual experiments
\cite{mpjr,yhua}. Such difficulties motivate people to construct
direct entanglement estimation scheme without any prior knowledge of the
quantum states and tomography. Recently, methods using
locally randomized measurements to probe entanglement have been put
forward \cite{abcf,tapb,akno,lkjw,arhr,snao,ypzl,jlas,hrjp,xsog}.
Among them, many efforts have been devoted to verify the positive
partial transpose (PPT) condition \cite{aper} based on the moments of
the randomized measurements \cite{arhr,ypzl,jlas,hrjp,xsog}.

Consider bipartite quantum states $\rho_{AB}$ in $H_A\otimes H_B$,
where $H_A$ and $H_B$ are $n$-dimensional Hilbert spaces. The well
known PPT criterion says that the partially transposed density
matrix $\rho_{AB}^{T_A}$ with respect to the subspace $H_A$ is
positive semidefinite if $\rho_{AB}$ is separable. Otherwise,
$\rho_{AB}$ must be entangled. Recently, it is found that the PPT
condition can also be illustrated by considering the so-called
partial transpose moments (PT moments),
$$
p_k\equiv Tr[(\rho_{AB}^{T_A})^k],~~~k=1,2,3\cdots,d,
$$
where $d=n^2$ is the dimension of the global
system $H_A\otimes H_B$.
The PT moments can be experimentally measured through global (bilocal) random unitary matrices
\cite{jlas,ypzl} or local randomized measurements \cite{arhr} based
on quantum shadow estimation \cite{hrjp}.
Obviously, $p_1=1$ for PPT states
such that $\rho_{AB}^{T_A}\geq 0$.
PT moments been first studied in
quantum field theory in quantifying the correlations in many-body
systems \cite{pjet}. In \cite{arhr}, the authors first show that the first
three PT moments can be used to define a simple yet powerful test
for bipartite entanglement: $p_3\geq p_2^2$ if $\rho_{AB}$ is a PPT state.
More recently, the authors in \cite{xsog}
obtain a family of separability criteria: a necessary
condition for $\rho_{AB}$ being a separable state is that
$B_{[(d-1)/2]}(\textbf{p}) \geq 0$, where
$[B_k(\textbf{p})]_{ij}=p_{i+j+1}$ for $i,j=0,1,2,\cdots,k$,
$\textbf{p}=(p_0,p_1,p_2,\cdots,p_d)$ with $p_0=d$.

In this paper, instead of the PT moments, we introduce rearrangement
moments based on the rearrangement operations \cite{kclw,orud}.
Similar to that given by the PT moments, we present quantum
separability criteria given by the rearrangement moments. By
detailed example, we show that the rearrangement moments based
criteria may detect entanglement.

\section{Separability criterion based on realignment moments}
In the following we present separability criteria based on the realignment of matrices,
by using results from the matrix analysis \cite{racr,racr1} and the
Hoelder's inequality.

For a given $m \times n$ matrix $C=[c_{ij}]$ with entries $c_{ij}$,
the vector $vec(C)$ is defined by
$$
vec(C)=(c_{11},\cdots,c_{m1},c_{12},\cdots,c_{m2},\cdots,c_{1n},\cdots,c_{mn}).
$$
Let $Z$ be an $m\times m$ block matrix with $n\times n$ block
matrices. The realigned matrix $R(Z)$ of $Z$ is defined by
\begin{eqnarray}\label{r1}R(Z)\equiv
\left(\begin{array}{ccccccc}
     vec(Z_{1,1}) \\
   \vdots \\
vec(Z_{m,1})\\
\vdots \\
vec(Z_{1,m})\\
\vdots \\
vec(Z_{m,m})
\end{array}\right).
\end{eqnarray}

Let $\rho_{AB}$ be a bipartite state in $H_A\otimes H_B$. We define the following quantities to be realignment moments of $\rho_{AB}$,
$$
r_k \equiv Tr[R(\rho_{AB})R(\rho_{AB})^{\dag}]^{\frac{k}{2}},~~~
k=1,2,3\cdots,d.
$$
For convenience we set $r_0=d$. We have the following conclusion on quantum separability in terms of the realignment moments.

\begin{theorem}
If the quantum state $\rho_{AB}$ is separable, then $r_2^2 \leq r_3$.
\end{theorem}

{\sf Proof}
Let $M$ be a Hermitian $d \times d$ matrix with eigenvalue
decomposition $M=\sum_{i=1}^d\lambda_i|x_i\rangle\langle x_i|$, where $\lambda_i$ and $|x_i\rangle$
are the eigenvalues and eigenvetors of $M$, respectively. The Schatten-p norm ($p \geq 1$) $\|M\|_{p}$ of $M$
is given by
$$
\|M\|_{p}\equiv
(\sum_{i=1}^d|\lambda_i|^p)^{1/p}=Tr(|M|^p)^{1/p},
$$
where
$|M|=\sqrt{MM^{\dag}}=\sqrt{M^2}=\sum_{i=1}^d|\lambda_i||x_i\rangle\langle
x_i|$. Set $M=(R(\rho_{AB})R(\rho_{AB})^{\dag})^{\frac{1}{2}}$. By
the definition of realignment moments, we obtain
$\|M\|^4_2=\|R(\rho_{AB})R(\rho_{AB})^{\dag}\|^4_2=r_2^2$,
$\|M\|_1=r_1$ and $\|M\|_3^3=r_3$. It has been shown in \cite{arhr}
that $\|M\|_{p}$ satisfies the following matrix norm inequality,
$\|M\|^4_2\leq \|M\|_1\|M\|_3^3$. That is for any state $\rho_{AB}$
we have $r_2^2 \leq r_1r_3$.

Moreover, in \cite{kclw} it has been shown that if a bipartite state
$\rho_{AB}$ is separable, then the trace norm $||R(\rho_{AB})||$ of
the realigned matrix $R(\rho_{AB})$, namely, the sum of all the
singular values of $R(\rho_{AB})$ satisfies $||R(\rho_{AB})||\equiv
Tr[R(\rho_{AB})R(\rho_{AB})^{\dag}]^{1/2}\leq 1$. Hence, if
$\rho_{AB}$ to be a separable state we have $r_1 \leq 1$, we can
draw the conclusion of the theorem.
\qed

Denote $\textbf{r}=(r_0,r_1,r_2,\cdots,r_d)$. We can construct the Hankel
matrices $[H_k(\textbf{r})]_{ij}=r_{i+j}$ and
$[B_k(\textbf{r})]_{ij}=r_{i+j+1}$ for $i,j=0,1,2,\cdots,k$ \cite{xsog}.
A stronger separability criterion involving higher order $r_k$ can be derived.

\begin{theorem} If $\rho_{AB}$ is separable, then $\widehat{H_{k}}(\textbf{r}) \geq 0$  for $k=1,2\cdots,\lfloor\frac{d}{2}\rfloor$ and
$\widehat{B_{l}}(\textbf{r}) \geq 0$ for $l=1,2,\cdots,
\lfloor\frac{d-1}{2}\rfloor$, where $\lfloor\cdot\rfloor$ is the
integer function, $\widehat{H_{k}}$ and $\widehat{B_{l}}$ are matrices
obtained by replacing $r_1$ with $1$ in $[H_k(\textbf{r})]$ and
$[B_k(\textbf{r})]$, respectively.
\end{theorem}

{\sf Proof} For any state $\rho_{AB}$
let $Y=(R(\rho_{AB})R(\rho_{AB})^{\dag})^{1/2}$. Denote
$\textbf{v}\equiv(v_0,v_1,\cdots,v_{\lfloor\frac{d}{2}\rfloor})=(I=Y^{0},Y,Y^2,\cdots,Y^{\lfloor\frac{d}{2}\rfloor})$
and
$\textbf{u}\equiv(u_0,u_1,\cdots,u_{\lfloor\frac{d-1}{2}\rfloor})=(Y^{\frac{1}{2}},Y^{\frac{3}{2}},\cdots,Y^{\lfloor\frac{d-1}{2}\rfloor})$.
Under the Hilbert-Schmidt inner product of matrices, the elements of
the Gram matrices for $\textbf{v}$ and $\textbf{u}$ are given by
$$
\langle v_i,v_j\rangle=Tr(Y^{i+j})=r_{i+j}
$$
and $$\langle
u_i,u_j\rangle=Tr(Y^{i+\frac{1}{2}}Y^{j+\frac{1}{2}})=Tr(Y^{i+j+1})=r_{i+j+1},$$
which are just the elements of the Hankel matrices $H_k(\textbf{r})$
and $B_k(\textbf{r})$. As the Gram matrices are always positive
semidefinite \cite{racr1}, we have $H_{k}(\textbf{r}) \geq 0$  for
$k=1,2\cdots,\lfloor\frac{d}{2}\rfloor$ and $B_{l}(\textbf{r}) \geq
0$ for $l=1,2,\cdots, \lfloor\frac{d-1}{2}\rfloor$ for any state.
Now, if $\rho_{AB}$ to be a separable state, we have $r_1 \leq 1$,
then let all $r_1$s in $\lfloor H_k(\textbf{r})\rfloor$ and $\lfloor
B_k(\textbf{r})\rfloor$ are replaced by $1$, we can draw the
conclusion of the theorem.
\qed

In particular, for $l=1$ in Theorem 2 one gets $B_1=\left(\begin{array}{cc}
    r_1 &  r_2 \\
    r_2 & r_3 \\
  \end{array}\right)$, then $r_1$ replaced by $1$, we get $\widehat{B_1}=\left(\begin{array}{cc}
    1 &  r_2 \\
    r_2 & r_3 \\
  \end{array}\right)\geq 0$, which gives rise to the conclusion of Theorem 1. Theorem 2 can be used
as effective separability criteria. It implies that if an
inequalities $\widehat{H_{k}(\textbf{r})} \geq 0$ for
$k=1,2\cdots,\lfloor \frac{d}{2}\rfloor$ or
$\widehat{B_{l}(\textbf{r})} \geq 0$ for $l=1,2,\cdots, \lfloor
\frac{d-1}{2}\rfloor$ is violated, $\rho_{AB}$ must be an entangled
state.

In fact, based on matrix realignment, we can also define the following moments,
$$
q_k \equiv Tr[R(\rho_{AB}-\rho_A\otimes
\rho_B)R(\rho_{AB}-\rho_A\otimes \rho_B)^{\dag}]^{k/2}
$$
for $ k=1,2,3\cdots,d.$ Set
$\textbf{q}=(q_0,q_1,\cdots,q_d)$, where $q_0=d$.
In a similar way we have the following results.

\begin{theorem} For any state $\rho_{AB}$, we have  $H_{k}(\textbf{q}) \geq 0$ for $k=1,2\cdots,\lfloor\frac{d}{2}\rfloor$ and
$B_{l}(\textbf{q}) \geq 0$ for $l=1,2,\cdots,
\lfloor\frac{d-1}{2}\rfloor$.
\end{theorem}

In \cite{zhcj,ogog} the authors show that if a bipartite state $\rho_{AB}$ is separable, then
$$
\|R(\rho_{AB}-\rho_A\otimes
\rho_B)\| \leq \sqrt{(1-Tr\rho_A^2)(1-Tr\rho_B^2)},
$$
where $\rho_A$ and $\rho_B$ are reduced matrices with respect to the subsystems $A$ and $B$.
Taking $l=1$ in Theorem 3 and using the above result, we have

\begin{corollary}
If $\rho_{AB}$ is a separable state, then
$q_2^2 \leq (1-Tr\rho_A^2)(1-Tr\rho_B^2)q_3$.
\end{corollary}

In the following we give a concrete example to illustrate our
entanglement criterion based on realignment moments. Let us
consider the Werner state,
$$
\rho_p=p|\psi\rangle\langle\psi|+\frac{1-p}{4}I_4,
$$
where $0 \leq p \leq 1$ and
$|\psi\rangle=\frac{1}{\sqrt{2}}(|00\rangle+|11\rangle)$, $I_4$ is
the $4 \times 4$ identity matrix.
By straightforward calculation we get $f\equiv r_3-r_2^2
=\frac{1}{16}(6p^3 + 1 - 9p^4 - 6p^2)$ as a function of $p$. The
inequality $r_2^2 \leq r_3$ in Theorem 1 is violated for $p>0.44$, see Fig. 1.
\begin{figure}[ht]
\centering
\includegraphics[width=3in]{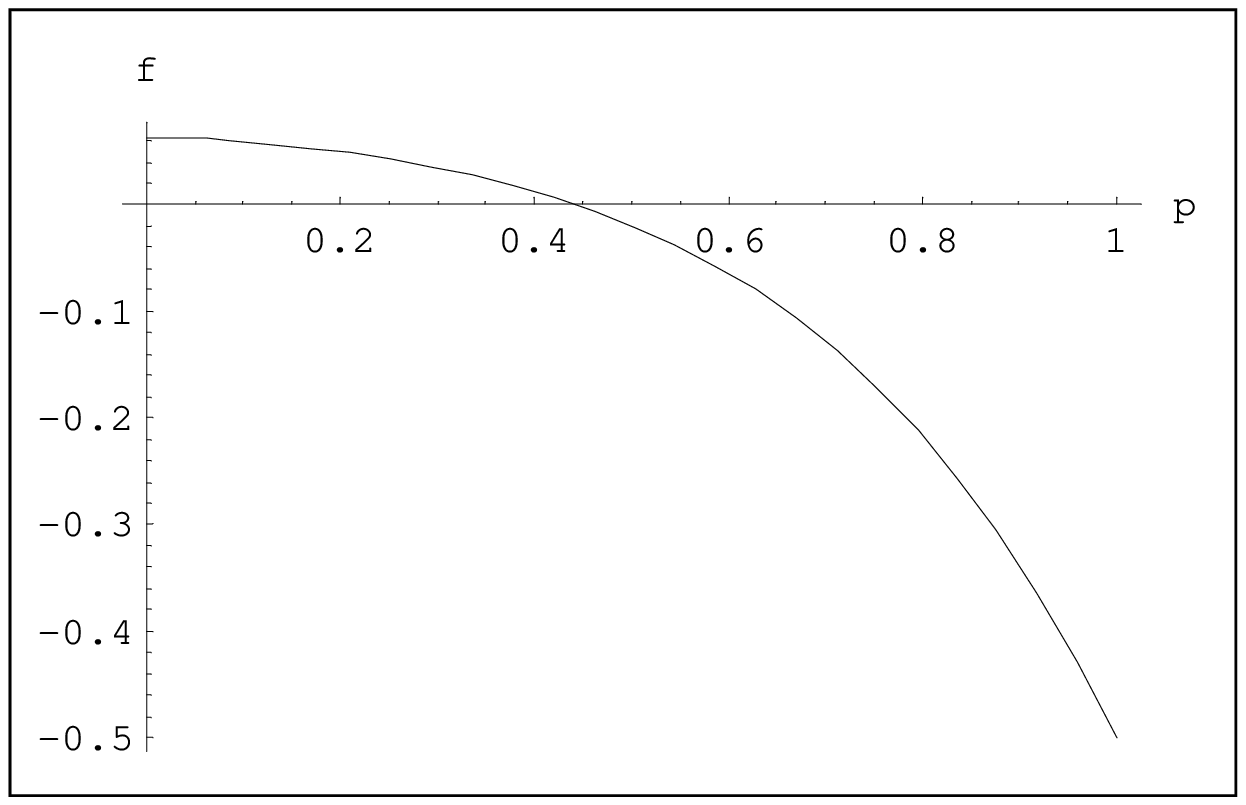}
\end{figure}

\section{Conclusions and discussions}
We have introduced two kinds of moments based on the realignment
operations. Based on these realignment moments we have obtained
entanglement criteria. Although we have not yet found an example
to show that the realignment moments based criteria work better than
other criteria like the PT moments based ones, the approach
may be used to to detect multipartite entanglement by using the generalized realignment
criteria \cite{lkcw,bjtm}. A scheme to experimentally measure the realignment moments via
measuring the purity $Tr(\rho_{AB}^2)$ \cite{abcf,tapb} or by global
random unitary matrices \cite{jlas,ypzl} or local randomized
measurements \cite{zlpz} would also be of significance.

\bigskip

Acknowledgments: This work is supported by the National Natural
Science Foundation of China under Grant Nos. 12126314, 12126351,
11861031, 12075159 and 12171044, Beijing Natural Science Foundation
(Z190005), the Hainan Provincial Natural Science Foundation of China
under Grant No.121RC539, and Academy for Multidisciplinary Studies,
Capital Normal University. This project is also supported by the
specific research fund of The Innovation Platform for Academicians
of Hainan Province under Grant No. YSPTZX202215.

\bigskip

Data availability statement: Our manuscript has no associated data.

\end{document}